\documentclass[aps,nofootinbib,groupedaddress]{revtex4}

\usepackage{amsmath,amssymb, amsthm,amstext}
\usepackage{natbib}
\usepackage{graphicx}
\usepackage{color}
\usepackage{array, enumerate}
\usepackage{bm}
\usepackage{multirow}
\usepackage[breaklinks,colorlinks,citecolor=blue]{hyperref}
\usepackage{pifont, ragged2e}

\newcommand{\cmark}{\ding{51}}%
\newcommand{\xmark}{\ding{55}}%

\def\be{\begin{equation}}
\def\ee{\end{equation}}
\newcommand{\DT}{\Delta}
\newcommand{\Dm}{\Delta_{\rm m}}

\newcommand{\Dg}{\Delta_{\rm g}}
\newcommand{\bne}{ \hat{\mathbf n}_e}
\newcommand{\bn}{ \hat{\mathbf n}}
\newcommand{\bre}{ {\mathbf r}_e}
\newcommand{\br}{ {\mathbf r}}
\newcommand{\V}{{\mathbf V}}
\newcommand{\bk}{\hat{\mathbf k}}
\newcommand{\Hm}{H_{\rm m}}
\newcommand{\rmv}{{\rm v}}
\newcommand{\rmq}{{\rm q}}

\begin{document}
\title{Forecasted constraints on modified gravity from Sunyaev Zel'dovich tomography}
\author{Zhen Pan}
\email{zpan@perimeterinstitute.ca}
\affiliation{Perimeter Institute for Theoretical Physics, Waterloo, Ontario N2L 2Y5, Canada}
\author{Matthew C. Johnson}
\email{mjohnson@perimeterinstitute.ca}
\affiliation{Department of Physics and Astronomy, York University, Toronto, Ontario, M3J 1P3, Canada}
\affiliation{Perimeter Institute for Theoretical Physics, Waterloo, Ontario N2L 2Y5, Canada}

\date{\today}
\begin{abstract}

Observational cosmology has become an important laboratory for testing General Relativity, with searches for modified gravity forming a significant portion of the science case for existing and future surveys. In this paper, we illustrate how future measurements of the Cosmic Microwave Background (CMB) temperature and polarization anisotropies can be combined with large galaxy surveys to improve constraints on modified gravity using the technique of Sunyaev Zel'dovich (SZ) tomography. SZ tomography uses the correlations between the kinetic/polarized SZ contributions to the small-angular scale CMB and the distribution of structure measured in a galaxy redshift survey to reconstruct the remote dipole and quadrupole fields, e.g. the CMB dipole and quadrupole observed throughout the Universe. We compute the effect of a class of modifications of gravity on the remote dipole and quadrupole fields, illustrating that these observables combine a number of the desirable features of existing probes. We then perform a fisher forecast of constraints on a two-parameter class of modifications of gravity for next-generation CMB experiments and galaxy surveys. By incorporating information from the reconstructed remote dipole and quadrupole fields, we find that it is possible to improve the constraints on this model by a factor of $\sim 2$ beyond what is possible with a galaxy survey alone.  We conclude that SZ tomography is a promising method for testing gravity with future cosmological datasets.

\end{abstract}
\maketitle

\section{Introduction}

Our observable Universe presents us with at least two distinct ways in which we can observationally constrain General Relativity (GR) in the strong gravity regime, where the full non-linear nature of Einstein's equations are manifest. The first is near the horizons of black holes, which have been probed by LIGO through the observation of gravitational radiation from binary black hole mergers~\cite{LIGOScientific:2018mvr} and the Event Horizon Telescope, which has imaged the vicinity of the supermassive black hole at the centre of M87~\cite{1435168}. These observations have placed important constraints on potential deviations from GR~\cite{TheLIGOScientific:2016src,Abbott:2018lct,LIGOScientific:2019fpa,1435177}, and promise to provide even more stringent tests in the future (e.g. \cite{Berti:2019xgr}). The second regime is on cosmological distance scales, of order the size of the observable Universe. The observed accelerated expansion of the Universe, which in the standard cosmological model $\Lambda$CDM is due to a cosmological constant, requires explanation in the strong gravity regime~\footnote{At an even more basic level, the full non-linear machinery of GR is necessary to understand how the averaged inhomogeneous distribution of matter on small scales induces expansion of the Universe on very large scales (see e.g.~\cite{Giblin:2016mjp}).}. In addition, the evolution (and observation) of inhomogeneities on ultra-large scales requires a fully relativistic treatment, and is sensitive to deviations from GR; see Refs.~\cite{Clifton:2011jh,Ishak:2018his} for recent reviews. Cosmic microwave background (CMB) experiments such as the Planck satellite~\cite{Ade:2015rim,Aghanim:2018eyx} and galaxy redshift surveys~\cite{Reyes:2010tr,Amon:2017lia,Joudaki:2017zdt,Abbott:2018xao,SpurioMancini:2019rxy} (through various combinations of clustering, redshift space distortions, and lensing) have put meaningful constraints on deviations from GR, and further tests are a primary science driver of future surveys such as Euclid~\cite{Laureijs2011} and LSST~\cite{0912.0201}.

In this paper, we explore the potential of kinetic Sunyaev Zel'dovich (kSZ) and polarized Sunyaev Zel'dovich (pSZ) tomography to test GR on cosmological scales. SZ tomography is used to denote the combination of kSZ and pSZ tomography. The kSZ effect~\cite{SZ80}, temperature anisotropies induced by the scattering of CMB photons from free electrons in bulk motion after reionization, is the dominant blackbody component of the CMB on small angular scales (corresponding to multipoles $\ell \agt 4000$). The observed kSZ temperature anisotropies from a given location in the observable Universe are determined by the product of the optical depth and the remote dipole field (e.g. the CMB dipole as observed from different points in spacetime) projected along the line of sight. The remote dipole field can be reconstructed from the correlations between a tracer of structure and the small-angular scale CMB using the technique of kSZ tomography~\cite{Ho09,Shao11b, Zhang11b, Zhang01,Munshi:2015anr,2016PhRvD..93h2002S,Ferraro:2016ymw,Hill:2016dta,Zhang10d,Zhang:2015uta,
Terrana2016,Yasini2016,Deutsch:2017ybc,Smith:2018bpn,Munchmeyer:2018eey,Sehgal:2019nmk}. The pSZ effect is the observed polarized component of CMB photons scattered after reionization, determined by the optical depth and the remote quadrupole field (the CMB quadrupole observed at different locations). The remote quadrupole field can be reconstructed analogously to the remote dipole field using pSZ tomography~\cite{Kamionkowski1997,Bunn2006,Portsmouth2004,2012PhRvD..85l3540A,Hall2014,Deutsch:2017cja,Deutsch:2017ybc,Louis:2017hoh,Meyers:2017rtf,Deutsch:2018umo}. Because kSZ/pSZ tomography reconstructs the remote dipole/quadrupole fields along our past light cone, these techniques can be used to probe the growth of structure over cosmic timescales, and therefore can in principle serve as a powerful probe of modifications to GR. A major limitation on using kSZ/pSZ tomography for measurement of growth is our inability to use a tracer of LSS to perfectly infer the distribution of electrons, a problem known as the optical depth degeneracy~\cite{Battaglia:2016xbi,Hall2014,Smith:2018bpn}. This can in principle be mitigated by measurements of other tracers of the electron distribution~\cite{Madhavacheril:2019buy}, and we discuss the impact of the optical depth degeneracy on our constraints below.

There are a number of theoretical approaches to modifying GR in the cosmological setting \cite{Mustapha2019}. One can take a top-down approach, constructing theoretically viable models at the level of the Lagrangian. Examples include $f(R)$ models~\cite{Hu:2007nk}, braneworld scenarios~\cite{Dvali:2000hr}, and Horndeski theories~\cite{Horndeski:1974wa}. An effective field theory (EFT) approach to cosmological perturbation theory~\cite{Cheung:2007st,Weinberg:2008hq} can incorporate modifications to GR systematically~\cite{Creminelli:2008wc,Park:2010cw,Bloomfield:2011np,Gubitosi:2012hu}, encoding the modifications as free coefficients/functions in the EFT. Parametric expansions about GR solutions, such as the Parameterized Post-Friedmann (PPF) framework~\cite{Hu2007, Baker:2012zs}, have also been employed. Both the EFT and PPF framework encode the effects of any top-down model consistent with various assumptions about symmetries and extra degrees of freedom. In this paper, we employ a phenomenological approach that directly modifies the linearized Einstein equations~\cite{Zhao:2008bn,PhysRevD.78.024015,2010PhRvD..81j4023P} by adding two free functions of time and scale, assuming that the background evolution remains as in $\Lambda$CDM. Our primary motivation for using this approach is that it can encode all modifications of GR in the scalar sector~\cite{Zhang:2007nk,Amendola:2007rr}, and it is employed in a variety of existing cosmological analyses (e.g.~\cite{Daniel:2010ky,Simpson:2012ra,Macaulay:2013swa,Aghanim:2018eyx,Ade:2015rim,Abbott:2018xao}) and Boltzmann codes~\cite{Hojjati:2011ix,Zhao:2008bn}. As our purpose is to illustrate the general utility of kSZ/pSZ tomography for constraining modifications of GR, we focus on the simplest possible parameterization in this class, where any scale dependence of modified growth is neglected and where the time-dependence is proportional to the fractional energy density in dark energy, assuming it is a cosmological constant. We do not undertake a comprehensive forecast of the constraints on more general models of modified gravity using the top-down, EFT, or PPF approaches.

Within this framework, the remote dipole and quadrupole fields are sensitive to modifications of GR in a number of ways. The largest contribution to the remote dipole field is from the local peculiar velocity field, which is sensitive to changes in the linear growth function. Both the remote dipole and quadrupole fields receive contributions from the integrated Sachs-Wolfe effect, which is sensitive to both the linear growth function and gravitational slip (e.g. the non-equality of the two Bardeen potentials). The remote dipole and quadrupole fields combine a number of desirable properties of other datasets used to constrain GR. The primary CMB is sensitive to modifications of GR primarily through the ISW effect, which is limited to a line of sight integral from our position to the surface of last scattering. However, the ISW contribution to the remote dipole/quadrupole fields can be measured at a number of redshifts, yielding more information about the evolution of growth and gravitational slip. The local Doppler contribution to the remote dipole field provides an excellent tomographic measurement of the line-of-sight peculiar velocity field, especially on the largest scales~\cite{Deutsch:2017ybc,Smith:2018bpn}, yielding a sensitive probe of growth. Just as the popular $E_G$ statistic~\cite{Zhang:2007nk} combines measurements of velocities and lensing to constrain both growth and gravitational slip while removing the degeneracy with galaxy bias, the full correlation structure of the remote dipole/quadrupole fields with a galaxy survey can be used to constrain modified GR while mitigating the effects of galaxy bias and the optical depth degeneracy.

Several previous works have investigated the utility of measurements of the kSZ effect for constraining proposed modifications of GR~\footnote{Several papers have also explored constraints on dark energy in the context of GR using the kSZ effect, e.g.~\cite{Bhattacharya:2007sk,Xu:2013jma,Ma:2013taq}}. Refs.~\cite{2009PhRvD..80f2003K,Mueller:2014nsa} forecasted constraints on a parameterized modification of the growth function from future CMB experiments using the pairwise velocity statistic. Formally, the information content of the pairwise velocity statistic should be equivalent to remote dipole reconstruction, as shown in~\cite{Smith:2018bpn}. A direct comparison with our work is not possible, since we choose to constrain a different class of models, however the degree of improvement offered by including kSZ measurements is roughly compatible with what is found here. Ref.~\cite{Bianchini:2015iaa} and \cite{Roncarelli2018} calculated the global kSZ power spectrum for a class of modifications of GR. This analysis did not take into account the information in cross-correlations with LSS measurements, and we therefore expect that the present analysis would provide more stringent constraints on such models. There are several advantages to the strategy employed here, as compared to previous work. First, remote dipole/quadrupole reconstruction isolates the relevant cosmological information content of the kSZ/pSZ effect: it is a powerful probe of {\em large-scale} inhomogeneities through cosmic time, and therefore the growth function in the linear regime. In addition, it is straightforward to calculate the covariances of the remote dipole/quadrupole fields with a variety of other cosmological probes such as the primary CMB temperature and polarization, galaxy number density, the lensing potential, etc. This provides a convenient framework in which to do joint forecasts, as we do here. Finally, the optical depth degeneracy can be incorporated simply as a redshift-dependent bias on the amplitude of the reconstructed fields which can be marginalized over, as shown in~\cite{Smith:2018bpn}.

The focus of the present work is to outline the impact of modifications of GR on the remote dipole and quadrupole fields, as well as to provide a forecast highlighting the improvement in constraints possible from kSZ/pSZ tomography using next-generation CMB experiments and LSS surveys. We find that significant improvement is possible, especially for gravitational slip, even in the scenario where galaxy bias and the optical depth degeneracy are fully marginalized over. The plan of the paper is as follows. In Section \ref{sec:mg}, we introduce the parameterization of modified gravity (MG) we use, detail the effect of MG
on the angular power spectra of galaxy number counts and remote dipole and quadrupole fields, and discuss the imprints of a preferred
reference frame associated with MG on the remote dipole field. In Section \ref{sec:fisher}, we perform a Fisher
forecast to explore how well the two free MG parameters we consider can be constrained using different datasets. We summarize our results in
Section \ref{sec:summary}. In Appendix \ref{sec:analytic} and \ref{sec:consistency} we also provide some analytic understanding of the effect of MG on growth in the large scale and small scale limits.

\section{Sunyaev Zel'dovich tomography in modified gravity}
\label{sec:mg}
In this section, we outline the phenomenological parameterization of MG that we consider and discuss the effect of such modifications on the observed galaxy number counts and the remote dipole/quadrupole fields. We consider scalar perturbations only, and assume that the expansion history is as in $\Lambda$CDM, using the best-fit cosmological parameters from the {\it Planck} temperature and lensing data \cite{Planck2015par}. The perturbed FRW metric in Newtonian gauge is written as
\be
    ds^2 = -(1+2\Psi(t, \vec x)) dt^2 + a^2(t)(1-2\Phi(t,\vec x)) d\vec x^2\ ,
\ee
where $\Psi$ and $\Phi$ are the Newtonian pentential and curvature perturbation, respectively. Within GR, in the absence of anisotropic stress, the linearized Einstein equations reduce to the Poisson equation and the condition $\Psi=\Phi$. Here, we consider parameterized modifications of GR characterized by two free functions of time and scale, $\mu$ and $\gamma$, defined by:
\begin{subequations}
  \begin{align}
    -k^2 \Psi &= 4\pi G a^2 (\rho_{\rm m} + \rho_{\rm r})\mu \DT  \ , \label{eq:Possion}\\
    \Phi/\Psi&= \gamma \ . \label{eq:slip}
  \end{align}
\end{subequations}
Here $\Delta$ is the total energy density perturbation defined by
$(\rho_{\rm m} + \rho_{\rm r})\DT = \rho_{\rm m} \Delta_{\rm m} + \rho_{\rm r} \Delta_{\rm r}$,
with $\Delta_{\rm m}$ and $\Delta_{\rm r}$ being the matter and radiation components, respectively. We note that
this definition of the total energy density perturbation assumes that any extra degrees of freedom associated with this class of
modifications of GR do not cluster. The parameter $\gamma$ is often referred to as gravitational ``slip".
This particular class of parameterized modifications of GR was introduced in Ref.~\cite{Zhao:2008bn}, has been incorporated into CAMB~\cite{Hojjati:2011ix,Zhao:2008bn},
and constrained using data from e.g. CFHTLenS~\cite{Simpson:2012ra}, Planck~\cite{Aghanim:2018eyx,Ade:2015rim}, and the Dark Energy Survey (DES)~\cite{Abbott:2018xao}.

In principle, both $\mu$ and $\gamma$ are functions of time and scale $k$. However, in this paper we focus on a simple model where
\be
  \begin{aligned}
      \mu    &= 1 + \delta \mu \times \Omega_{\rm DE}(a) \ , \\
      \gamma &= 1 + \delta \gamma\times \Omega_{\rm DE}(a)\ ,
  \end{aligned}
\ee
 where $\Omega_{\rm DE}(a)$ is the dark energy fraction of the total energy density assuming $\Lambda$CDM and
 $ \delta \mu$, $\delta \gamma$ are constants. This choice of time dependence is motivated by the potential relation between
 modifications of GR and the observed accelerated expansion of the Universe. This model corresponds to the ``DE Related" parameterization
from Planck~\cite{Aghanim:2018eyx,Ade:2015rim}; mapping between variables we have $\delta \mu = E_{11}$ and $\delta \gamma = E_{22}$.
This is by no means a comprehensive analysis, but is meant to give a flavor of what kSZ and pSZ tomography could add to
the study of modified gravity on cosmological scales.

In the following subsections, we outline the effect of this parameterized modification of GR on the observables relevant for kSZ/pSZ tomography,
including the observed galaxy number counts and the remote dipole and quadrupole fields.

\subsection{Galaxy number counts in modified gravity}

The overdensity of galaxy number counts is written as \cite[e.g.,][]{Dio2013,Lorenz2018}
\be
\Dg(\hat{\mathbf n}, z) = b_{\rm g}(z) \Dm(\hat{\mathbf n}, z)  + \frac{1}{\mathcal H(z)} \partial_r(\mathbf V(\hat{\mathbf n}, z)  \cdot \hat{\mathbf n} ) + \frac{2-5s(z)}{2\chi(z)}\int_0^{\chi(z)} d\chi \left(2- \frac{\chi(z)-\chi}{\chi}\Delta_\Omega \right) (\Psi+\Phi)\ ,
\ee
where the second and third terms are the contributions from redshift space distortion and from gravitational lensing, respectively.
Here $b_{\rm g}$ is the galaxy bias, $\Dm$ is the matter overdensity, $\V$ is the matter velocity field, and $\mathcal H(z) = H(z) a$ is the comoving Hubble parameter. In the lensing term, $\chi(z) = \int_0^{z} dz/H(z)$ is comoving radial coordinate, $\Delta_\Omega$ is the angular Laplacian, and $s$ is the so-called magnification bias characterizing the galaxy number density
dependence on the luminosity and we calculate its fiducial values following the treatment of Ref.~\cite{Contreras:2019bxy}. In the following, we assume a fiducial galaxy bias model given by \cite{0912.0201}
\begin{equation}
b_{\rm g}(z) = 0.95/D(z)\ ,
\end{equation}
where $D(z)$ is the growth factor normalised as $D(z=0)=1$.
We neglect the relativistic corrections to the observed galaxy number counts~\cite{Yoo:2008tj,2009PhRvD..80h3514Y,Yoo:2010ni,1105.5280,1105.5292,1107.5427}. While kSZ tomography can be used to measure relativistic corrections~\cite{Contreras:2019bxy}, the effects from the parameterized modifications of GR considered here are expected to be unobservably small~\cite{Baker:2015bva}. We have confirmed that including relativistic corrections into the forecast below does not significantly affect our results.

\begin{figure}
\includegraphics[scale=0.55]{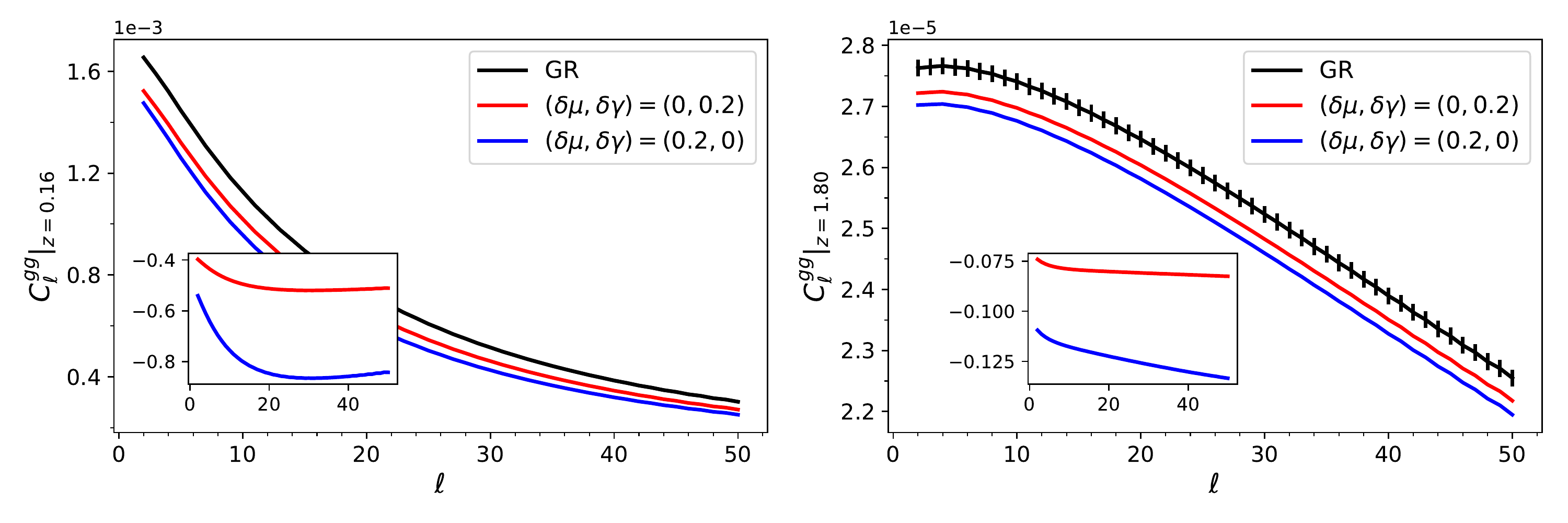}
\caption{\label{fig:clgg} The dependence of the galaxy number counts power spectrum $C_\ell^{\rm gg}$ on the two MG parameters. The error bars shown are
calculated assuming the shot noise of the LSST gold sample \cite{0912.0201} (in the left panel, the error bars are too small to
stick out), and the inset plots are the corresponding fractional change $\frac{dC_\ell^{\rm gg}}{d\theta}/C_\ell^{\rm gg}$,
with $\theta$ being $\delta\mu$ or $\delta\gamma$.}
\end{figure}

In Fig.~\ref{fig:clgg}, we show the dependence of the galaxy angular power
spectrum $C_\ell^{\rm gg}$ on the two MG parameters, where
evolution of matter perturbation $\Dm(a,k)$, velocity $V(a,k)$ and potential $(\Psi+\Phi)(a,k)$ are calculated using MGCAMB \cite{Zhao:2008bn,Hojjati:2011ix}. From the plot, we can see  $\delta\gamma$
induces a nearly scale-independent fractional change to $C_\ell^{\rm gg}$, which makes its effect on $C_\ell^{\rm gg}$ highly degenerate with galaxy bias $b_g(z)$; $\delta\mu$ induces a larger, scale-dependent
fractional change. Especially, we find the lensing term play an import role in breaking the degeneracy via its sensitivity
on the potential change in MG (see Appendix \ref{sec:analytic}).

\subsection{The remote dipole and quadrupole fields in modified gravity}
The kSZ effect is the result of Compton scattering of CMB photons by free electrons moving with respect
to the CMB rest frame. This produces CMB temperature anisotropies given by
\be
\left. \frac { \Delta T } { T } \right| _ { \mathrm { kSZ } } \left( \hat { \mathbf { n } }_ { e }  \right) =  \int d \chi _ { e } \ \dot{\tau} \left( \hat { \mathbf { n } } _ { e } , \chi _ { e } \right) v _ { \mathrm { eff } } \left( \hat { \mathbf { n } } _ { e } , \chi _ { e } \right)
\ee
where $\chi_e = \int_0^{z_e} dz/H(z)$ is comoving radial coordinate to the electron along the past light cone and $\bne$ is the angular
direction on the sky to the electron. The differential optical depth is defined as
\begin{equation}
\dot{\tau} \left( \hat { \mathbf { n } } _ { e } , \chi _ { e } \right) \equiv - \sigma_T a(\chi_e) \bar{n}_e (\chi_e) \left[ 1 + \delta_e \left( \hat { \mathbf { n } } _ { e } , \chi _ { e } \right) \right]
\end{equation}
where $\sigma_T$ is the Thomson scattering cross section, $\bar{n}_e (\chi_e)$ is the mean electron number density, $\delta_e \left( \hat { \mathbf { n } } _ { e } , \chi _ { e } \right)$ is the electron overdensity field. The remote dipole field $v _ { \mathrm { eff } } \left( \hat { \mathbf { n } } _ { e } , \chi _ { e } \right)$ observed by each electron projected along the line of sight is
\be
v _ { \mathrm { eff } } \left( \hat { \mathbf { n } } _ { e } , \chi _ { e } \right) \equiv \sum_{m=-1}^{1} \Theta_{1}^m \left( \hat { \mathbf { n } } _ { e } , \chi _ { e } \right) Y_{1m} \left(\hat { \mathbf { n } } _ { e } \right), \ \ \ \ \Theta_{1}^m \left( \hat { \mathbf { n } } _ { e } , \chi _ { e } \right) \equiv \int \ d^2 \bn \Theta(\bne, \chi_e, \bn) Y_{1m}(\bn)
\ee
where $\Theta(\bne, \chi_e, \bn)$ is the CMB temperature the electron sees along direction $\bn$,
$\Theta(\bne, \chi_e, \bn) = \Theta_{\rm SW} + \Theta_{\rm ISW} + \Theta_{\rm Dop} $. The SW contribution is given by
\be
\begin{aligned}
  \label{eq:sw}
  \Theta_{\rm SW}(\bne, \chi_e, \bn)
  = \frac{1}{3}\Psi(\chi_{\rm dec}, \mathbf{r}_{\rm dec})
  = \frac{1}{3} D_\Psi(\chi_{\rm dec}, \mathbf{r}_{\rm dec}) \Psi_i(\mathbf{r}_{\rm dec})\ ,
\end{aligned}
\ee
where $\mathbf{r}_{\rm dec} = \chi_e\bne + \chi_{\rm dec}^e \bn$ with $\chi_{\rm dec}^e = \chi_{\rm dec}-\chi_e$,
and $D_\Psi$ is the potential growth function defined by $\Psi(a, {\mathbf r}) = D_\Psi(a, {\mathbf r})\Psi_i({\mathbf r})$.
We have used the fact that the MG effect is negligible at early times, i.e., $\gamma_i = \gamma_{\rm dec} = 1$.
The ISW contribution is given by
\be
\begin{aligned}
  \label{eq:isw}
  \Theta_{\rm ISW}(\bne, \chi_e, \bn)
   = \int_{a_{\rm dec}}^a \frac{d(\Psi+\Phi)}{da} da
   = \int_{a_{\rm dec}}^a \frac{d(1+\gamma)D_\Psi}{da}  \Psi_i da  .
\end{aligned}
\ee
The Doppler contribution is given by the relative velocity of the emitter and the scatter,
\be
\label{eq:dop}
\Theta_{\rm Dop}(\bne, \chi_e, \bn) = \bn \cdot[\V(\bre, \chi_e)-\V(\br_{\rm dec}, \chi_{\rm dec})] \ .
\ee
Similar to the potential growth function $D_\Psi$, we can define a velocity growth function $D_v(a,k)$
connecting the velocity to the primordial potential perturbation by
$\mathbf{V} = - D_v(a, \mathbf{r}) \nabla \Psi_i(\mathbf{r})$.

\begin{figure}
\includegraphics[scale=0.55]{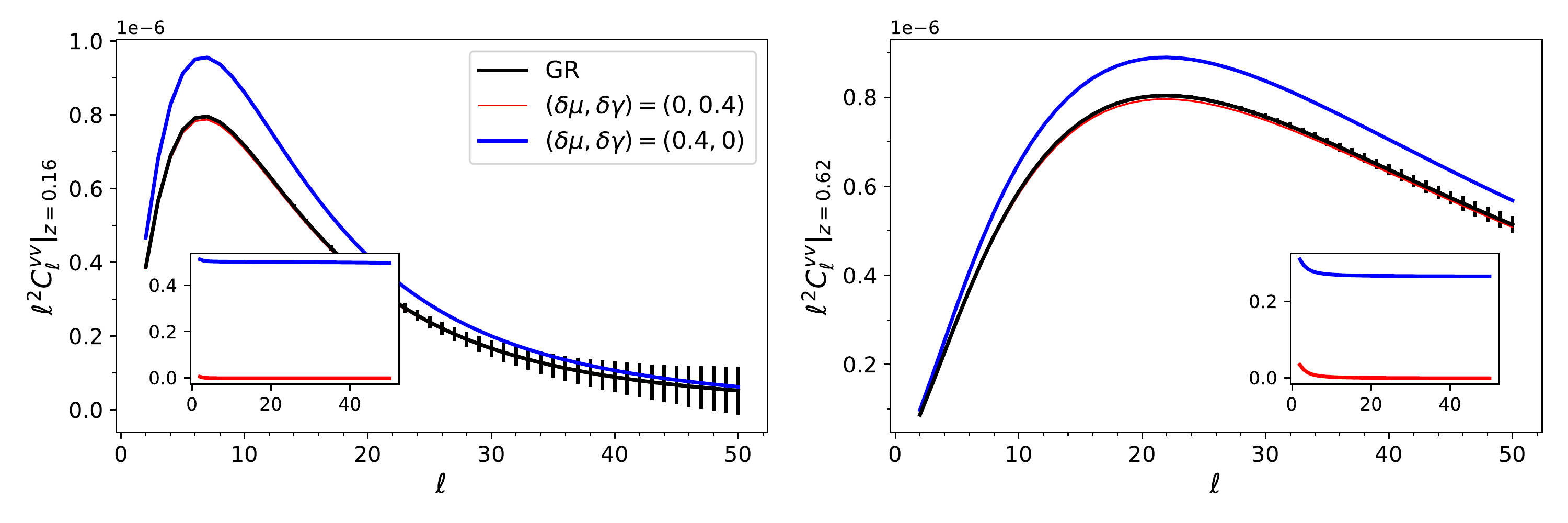}
\includegraphics[scale=0.55]{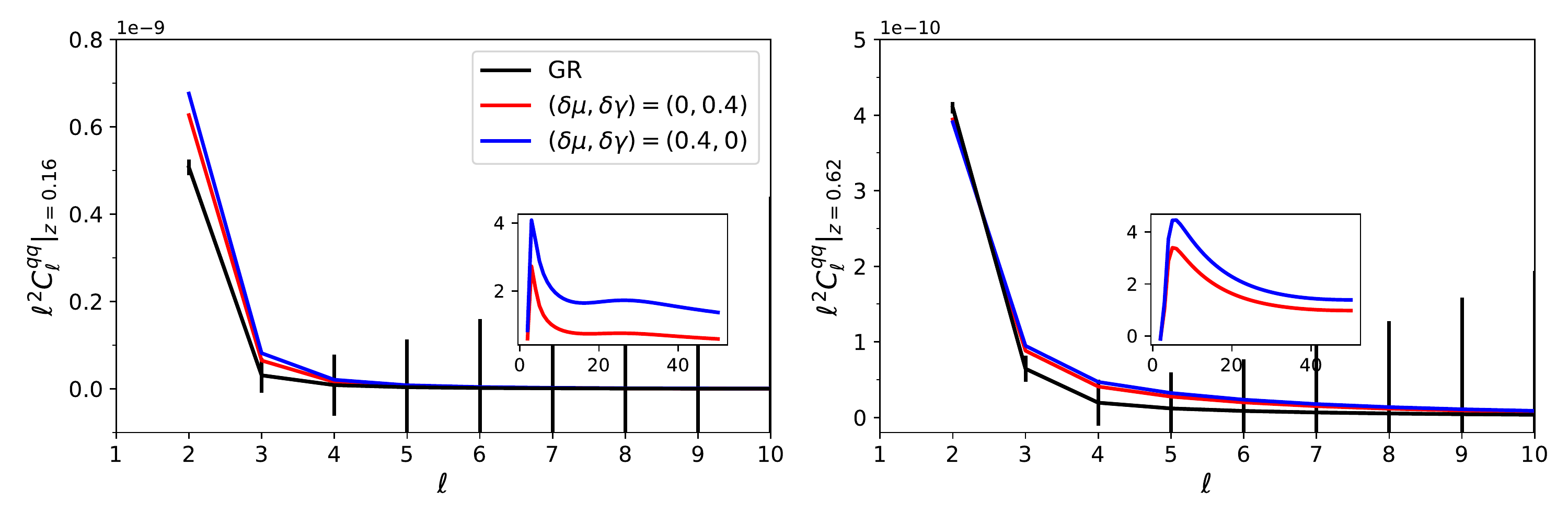}
\caption{\label{fig:clqv} The effect of MG on the remote dipole and quadrupole signals ($C_\ell^{\rm vv}$ and $C_\ell^{\rm qq}$),
where the inset plots are the corresponding fractional change $\frac{dC_\ell^{\rm gg}}{d\theta}/C_\ell^{\rm gg}$,
with $\theta$ being $\delta\mu$ or $\delta\gamma$. The error bars are the corresponding
kSZ and pSZ reconstruction noise expected from
cross correlating a LSS survey with a CMB experiment detailed in Sec.~\ref{sec:fisher}. }
\end{figure}

We can define a kernel in Fourier space relating the primordial
potential $\Psi_i(k)$ to the remote dipole field~\cite{Terrana2016},
\be
v _ { \mathrm { eff } } \left( \hat { \mathbf { n } } _ { e } , \chi _ { e } \right) = i\int \frac{d^3k}{(2\pi)^2} \Psi_i(\mathbf k)\mathcal K(k,\chi_e)
\mathcal P_1(\bk \cdot \bne) e^{i\chi_e {\mathbf k}\cdot \bne}
\ee
$\mathcal P_n$ is the Legendre polynomial of degree $n$.
The Fourier kernel is $\mathcal K(k, \chi_e) = \mathcal K_{\rm SW}(k, \chi_e)+ \mathcal K_{\rm ISW}(k, \chi_e)+ \mathcal K_{\rm Dop}(k, \chi_e)$ with
each component given by
\be\label{eq:vkernel}
\begin{aligned}
  \mathcal K_{\rm SW}(k, \chi_e)
  &= 3 \left(\frac{1}{3}D_\Psi(k, \chi_{\rm dec}) \right) j_1(k\chi_{\rm dec}^e)\ , \\
  \mathcal K_{\rm ISW}(k, \chi_e)
   &= 3 \int_{a_{\rm dec}}^{a_e} \frac{d(1+\gamma)D_\Psi(k, \chi_a)}{da}  j_1(k\chi_a^e) da \ , \\
  \mathcal K_{\rm Dop}(k,\chi_e)
  & = kD_v(k, \chi_{\rm dec})\left[j_0(k \chi_{\rm dec}^e) -2 j_2(k \chi_{\rm dec}^e)\right] -k D_v(k, \chi_e) \ ,
\end{aligned}
\ee
where the growth functions $D_\Psi(a,k)$ and $D_v(a,k)$ are computed using MGCAMB \cite{Zhao:2008bn,Hojjati:2011ix}.
From Eq.~(\ref{eq:vkernel}), we conclude: $\mathcal K_{\rm SW}$ is determined by physics at the time of
recombination, and is therefore insensitive to the modifications of gravity we consider here; MG affects
 $\mathcal K_{\rm ISW}$ and $\mathcal K_{\rm Dop}$; because the Doppler term is the dominant contribution to the
 remote dipole field, most of the constraining power on MG comes from this term.
In top two panels of Fig.~{\ref{fig:clqv}}, we show the dependence of the angular power spectrum of the remote dipole field $C_\ell^{\rm vv}$ on the two
MG parameters. We see that $C_\ell^{\rm vv}$ is only sensitive to $\delta\mu$, not $\delta\gamma$. The reason is that
$C_\ell^{\rm vv}$ is mainly sourced by $\mathcal K_{\rm Dop}$ on sub-horizon scales, and $D_v(a,k)$ only depends
on $\delta\mu$, not $\delta\gamma$ (see Appendix \ref{sec:analytic}).

The pSZ effect is the polarized component of scattered photons after reionization, and the perturbations to the Stokes parameters are given by:
\begin{equation}
( Q \pm i U ) ^ { \mathrm { pSZ } } \left( \hat { \mathbf { n } } _ { e } \right) = \frac { \sqrt { 6 } } { 10 } \int d \chi _ { e } \ \dot{\tau} \left( \hat { \mathbf { n } } _ { e } , \chi _ { e } \right) q _ { \mathrm { eff } } ^ { \pm } \left( \hat { \mathbf { n } } _ { e } , \chi _ { e } \right),
\end{equation}
where $q _ { \mathrm { eff } } ^ { \pm } \left( \hat { \mathbf { n } } _ { e } , \chi _ { e } \right)$ is the remote quadrupole field, which receives contributions from both scalar and tensor modes; we consider only scalar contributions here, for a detailed discussion of the tensor contribution see Refs.~\cite{2012PhRvD..85l3540A,Deutsch:2017ybc,Deutsch:2018umo}. For scalar modes,
\begin{equation}
q _ { \mathrm { eff } } ^ {\pm} \left( \hat { \mathbf { n } } _ { e } , \chi _ { e } \right) = q _ { \mathrm { eff } } ^ {E} \left( \hat { \mathbf { n } } _ { e } , \chi _ { e } \right) \equiv \sum_{m=-2}^2 \Theta_{2}^m \left( \hat { \mathbf { n } } _ { e } , \chi _ { e } \right) _{\pm 2} Y_{2m} \left(\hat { \mathbf { n } } _ { e } \right)
\end{equation}
where $q _ { \mathrm { eff } } ^ {E}$ is an E-mode type remote quadrupole~\cite{2012PhRvD..85l3540A,Deutsch:2017ybc,Deutsch:2018umo}. In analogy to the remote dipole, we can relate the remote quadrupole to the primordial potential $\Psi_i(\bf k)$ by \cite{Deutsch:2017ybc}
\be
\Theta_{2}^m (\bne, \chi_e) =  \int \ d^2 \bn \ \Theta(\bne, \chi_e, \bn) Y_{2m}(\bn) = \int \frac{d^3k}{(2\pi)^2} \Psi_i(\mathbf k)\mathcal G(k,\chi_e)
 Y^*_{2m}(\bk \cdot \bne) e^{i\chi_e {\mathbf k}\cdot \bne}\ ,
\ee
where $\mathcal G(k,\chi_e) = \mathcal G_{\rm SW}(k,\chi_e) + \mathcal G_{\rm ISW}(k,\chi_e) + \mathcal G_{\rm Dop}(k,\chi_e)$ with each component given by
\be\label{eq:qkernel}
\begin{aligned}
  \mathcal G_{\rm SW}(k, \chi_e)&= -4\pi \left(\frac{1}{3}D_\Psi(k, \chi_{\rm dec})\right) j_2(k\chi_{\rm dec}^e)\ , \\
    \mathcal G_{\rm ISW}(k, \chi_e) &= -4\pi \int_{a_{\rm dec}}^a  \frac{d(D_\Psi(1+\gamma))}{da}  j_2(k\chi_a^e) da\ , \\
      \mathcal G_{\rm Dop}(k, \chi_e) &= \frac{4\pi}{5}kD_v(k,\chi_{\rm dec}) \left[3j_3(k\chi_{\rm dec}^e)
      -2j_1(k\chi_{\rm dec}^e) \right]\ .
\end{aligned}
\ee
In bottom two panels of Fig.~{\ref{fig:clqv}}, we show the dependence of the E-mode remote quadrupole $C_\ell^{\rm qq}$ on the two
MG parameters. Here we see that $C_\ell^{\rm qq}$ is sensitive to both MG parameters $\delta\mu$ and $\delta\gamma$ since the
ISW contribution is more important than for the remote dipole field.

The remote dipole and quadrupole fields defined above can be reconstructed from maps of the CMB temperature/polarization anisotropies and a three-dimensional probe of structure such as a galaxy redshift survey using the technique of SZ tomography. The galaxy survey is used to trace the differential optical depth at different locations, assuming a model for the correlation between electron and galaxy overdensity. Binning the galaxies in redshift, an unbiased quadratic estimator can be defined to give a three-dimensional reconstruction of the remote dipole and quadrupole fields. These are the primary observables which we employ below to obtain constraints on MG. The reconstruction noises are given in Ref.~\cite{Deutsch:2017cja}, to which we refer the reader for more details. Generally speaking, the reconstruction noise decreases with decreasing shot noise in the galaxy and with increasing sensitivity and resolution of the CMB experiment.

\subsection{The tilted Universe in modified gravity}

For adiabatic perturbations in GR, a pure gradient curvature perturbation can be removed by a coordinate transformation (see e.g.~\cite{Weinberg2003,Pajer:2017hmb}). This implies that, in the absence of a preferred reference frame, gradient modes can have no observable consequences. This so-called ``tilted Universe"~\cite{Turner91} was analyzed in detail by~\cite{Erickcek08} (see also Ref.~\cite{Mirbabayi:2014hda}), who showed that in Newtonian gauge there is a precise cancellation in the CMB temperature anisotropies between the SW, ISW, and Doppler contributions which ensure that a pure gradient mode is unobservable. The kSZ signal was shown to vanish in a tilted Universe in Refs.~\cite{Zhang:2015uta,Terrana2016}, which relies on a more stringent cancellation which must occur {\em everywhere} in the post-recombination Universe. In particular, the following relation between growth functions must hold in Newtonian gauge:
\begin{equation}\label{eq:GR_growth_relation}
F(a)\equiv\left( \frac{1}{3} D_{\Psi} (a_{\rm dec})\right) (\chi(a_{\rm dec}) - \chi(a)) - D_v(a) + D_v (a_{\rm dec}) +  \int_{a_{\rm dec}}^a da' \frac{d (1+\gamma)D_{\Psi}}{da'} (\chi(a_{\rm dec}) - \chi(a')) = 0,
\end{equation}
i.e., $\lim_{k\rightarrow0} \partial \mathcal K(k,a)/\partial k = 0$,
which was shown to hold within $\Lambda$CDM in Ref.~\cite{Terrana2016}.

\begin{figure}
\includegraphics[scale=0.55]{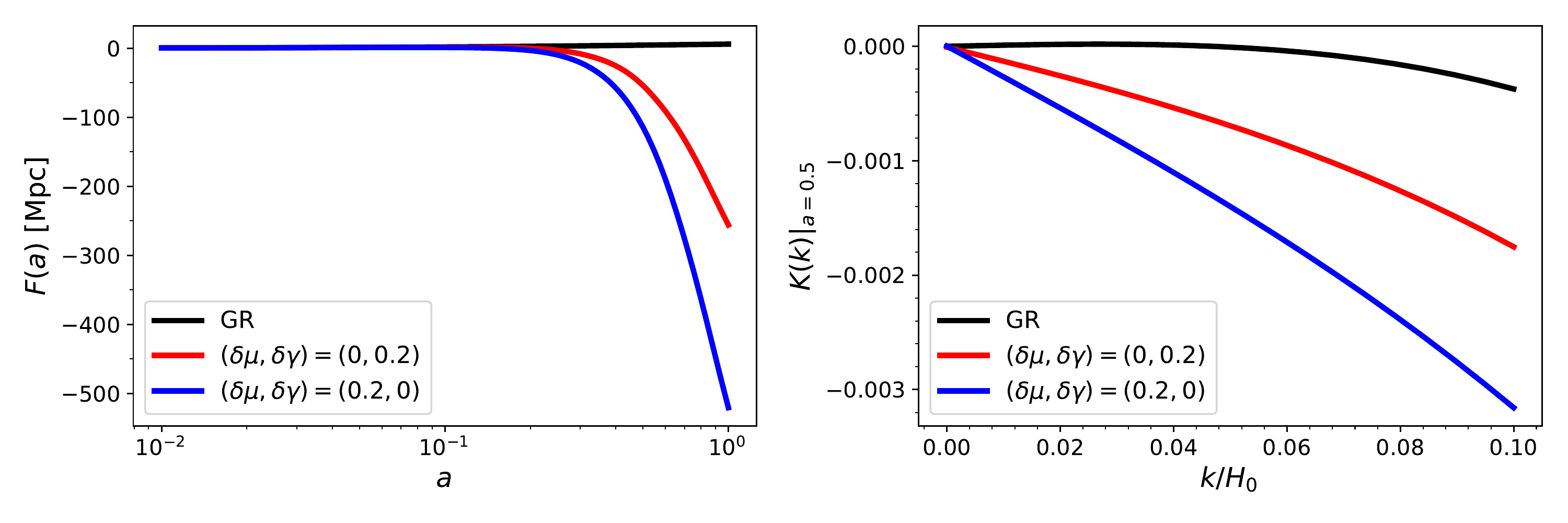}
\caption{\label{fig:v_ker}
In the left panel, we show function $F(a) =\lim_{k\rightarrow0} \partial\mathcal K(a,k)/\partial k$ (see Eq.~\ref{eq:GR_growth_relation}) in GR and MG as functions of scale factor $a$.
In the right panel, we show the kernel $\mathcal K(k)$ at $a=0.5$ for superhorizon modes,
where $\mathcal K(k)|_{\rm GR}\propto k^3$ while $\mathcal K(k)|_{\rm MG}\propto k$ in the small $k$ limit. This demonstrates that there are observable imprints of a preferred frame on the remote dipole field in MG.}
\end{figure}

Modifications of GR that alter growth will generically violate Eq.~(\ref{eq:GR_growth_relation}), thus making a gradient mode observable via the primary CMB and the kSZ effect. We plot Eq.~(\ref{eq:GR_growth_relation}) in the left panel of Fig.~\ref{fig:v_ker} for non-zero $\delta \mu$ and $\delta \gamma$ to illustrate this. Because we assume that modified growth does not occur until the onset of dark energy domination, we can see that Eq.~(\ref{eq:GR_growth_relation}) is violated only at late times. The contribution of a gradient mode to the remote dipole field is determined by the first derivative of the Fourier kernel, Eq.~(\ref{eq:vkernel}), in the limit where $k \rightarrow 0$. In GR, this is zero, implying that the leading-order behavior is $\mathcal K(k)|_{\rm GR}\propto k^3$. In Fig.~\ref{fig:v_ker}, we show the remote dipole Fourier kernel for GR as well as for non-zero $\delta \mu$ and $\delta \gamma$. Here, we see that for a generic non-zero $\delta \mu$ and $\delta \gamma$, the leading order behavior is $\mathcal K^v(k)|_{\rm GR}\propto k$. One consequence of this is that the remote dipole field within MG can be relatively more sensitive to super-horizon perturbations than in GR. However the kSZ signal is mainly sourced by the sub-horizon Doppler contribution,
and therefore the extra sensitivity to super-horizon perturbations in MG is hard to observe.

The in-principle observability of superhorizon gradient modes implies the existence of a preferred reference frame. In the parameterized modification of GR that we consider in this paper, the introduction of time-dependent functions $\mu$ and $\gamma$ explicitly break diffeomorphism invariance, and fix the preferred frame. In a top-down approach, this could correspond to an explicit or implicit breaking of diffeomorphism invariance due to the existence of additional degrees of freedom in the gravitational sector. Because the preferred frame is manifest only at late times, in the absence of more theoretical input, there is no reason that the frame setting the initial conditions for perturbations in the early Universe should be equivalent to the preferred frame fixed at late times. One could incorporate this uncertainty by introducing three extra parameters, corresponding to the components of a pure-gradient mode in $\Psi$. Here, we simply note that growth can be a probe of diffeomorphism invariance via Eq.~(\ref{eq:GR_growth_relation}), and that the remote dipole field can therefore be a unique probe of diffeomorphism invariance in theories that modify GR.

\section{Fisher Forecast}
\label{sec:fisher}
With some intuition about the effect of MG on galaxy number counts and the remote dipole/quadrupole field, we
now perform a Fisher forecast of the constraints on $\delta\mu$ and $\delta\gamma$ from forthcoming
CMB experiments and galaxy surveys using SZ tomography.

\subsection{Experiments}

The remote dipole and quadrupole fields can be reconstructed by cross correlating large-scale structure (LSS)
with small-scale CMB anisotropies using the quadratic estimators defined in~\cite{Deutsch:2017cja}.
The reconstruction noise on the remote dipole and quadrupole fields depends on the volume
and shot noise of the galaxy survey and the sensitivity the CMB experiment; we refer the reader to Ref.~\cite{Deutsch:2017cja} for
further details. We assume data on the full sky, and neglect foregrounds and systematics in both the CMB experiment and
galaxy survey. These are clearly idealistic assumptions, but should give a flavor of what is in-principle
obtainable with future experiments.

In this paper, we use the LSST gold sample as our fiducial LSS experiment.
For this dataset, the galaxy number density $n(z)$ per square arcmin is expected to be \cite{0912.0201}
\be
n(z) = n_{\rm g} \frac{1}{2z_0} \left( \frac{z}{z_0}\right)^2 \exp\left(\frac{z}{z_0} \right)\ ,
\ee
with $z_0 = 0.3$ and $n_{\rm g}=40/{\rm arcmin}^2$. The predicted photo-z error is $\sigma_z = 0.03(1+z)$,
which determines the minimum width of our redshift bins.

We consider hypothetical CMB experiments with a beam full-width-half-maximum (FWHM) $\theta_{\rm FWHM}$
of 1.5 arcmins and effective
detector noise level $\Delta_{\rm T} = \{5.0, 1.0, 0.1\}$ $\mu$K-armin, i.e.,
$N_\ell^{\rm TT} = \Delta_{\rm T}^2 \exp\left(\ell(\ell+1)\theta_{\rm FWHM}^2/8\ln 2 \right)$
and $N_\ell^{\rm EE} = 2N_\ell^{\rm TT}$. The expected noise $N_\ell^{ \{\rm vq\}\{\rm vq\} } $
of reconstructed kSZ/pSZ signals is calculated following Ref.~\cite{Smith:2018bpn} and
we show the noise level expected from LSST gold sample and the CMB experiment with optimal sensitivity ($\Delta_{\rm T}=0.1$ $\mu$K-armin) in Fig.~\ref{fig:clqv}.

\subsection{Forecasts}

The Fisher matrix for model parameters $\theta$ constrained by angular power spectra $C_\ell$ is written as \cite[e.g.,][]{Dodelson2003}
\be
F_{\alpha\beta} = \sum_\ell^{\ell_{\rm max}} \frac{2\ell+1}{2} f_{\rm sky}
{\rm Tr}\left(C_\ell^{-1} \frac{\partial C_\ell}{\partial \theta_\alpha} C_\ell^{-1}
\frac{\partial C_\ell}{\partial \theta_\beta}  \right)\ ,
\ee
and it is related to the expected uncertainty of a model parameter $\theta_\alpha$ by
\be
\sigma(\theta_\alpha) = \sqrt{(F^{-1})_{\alpha\alpha}}\ ,
\ee
where $f_{\rm sky}$ is the (mutual) sky fraction covered by the surveys, $C_\ell = C_\ell^{\rm XY} + N_\ell^{\rm XY}$
with $C_\ell^{\rm XY}$ and $N_\ell^{\rm XY}$ being the cross spectra of signals and noises, respectively;
${\rm X}$ and ${\rm Y}$ are the corresponding observables. To model the angular power spectra of galaxy number counts $C_\ell^{{\rm g}_i {\rm g}_j}$ across all redshift bins $[z_i, z_{i+1}]\times [z_j, z_{j+1}]$  ($i,j=1, 2, ..., N_{\rm bins}$),
we need $ 2N_{\rm bins}+2$ parameters $\{b_{\rm g}^i, s^i, \delta\mu, \delta\gamma\}$ (all $\Lambda$CDM parameters
are assumed fixed). Due to the optical depth degeneracy \cite{Battaglia:2016xbi,Hall2014,Smith:2018bpn}, the remote dipole and quadrupole fields reconstructed from kSZ/pSZ tomography are uncertain up to an optical depth bias $b_v^i$ in each redshift bin. To model the angular power spectra of the dipole/quadrupole fields $C_\ell^{\{\rmv_i\rmq_m\} \{\rmv_j\rmq_n\}}$ across all redshift bins ($i,j, m,n=1, 2, ..., N_{\rm bins}$),
we need $ N_{\rm bins}+2$ parameters $\{b_v^i, \delta\mu, \delta\gamma\}$.

\begin{figure}
\includegraphics[scale=0.55]{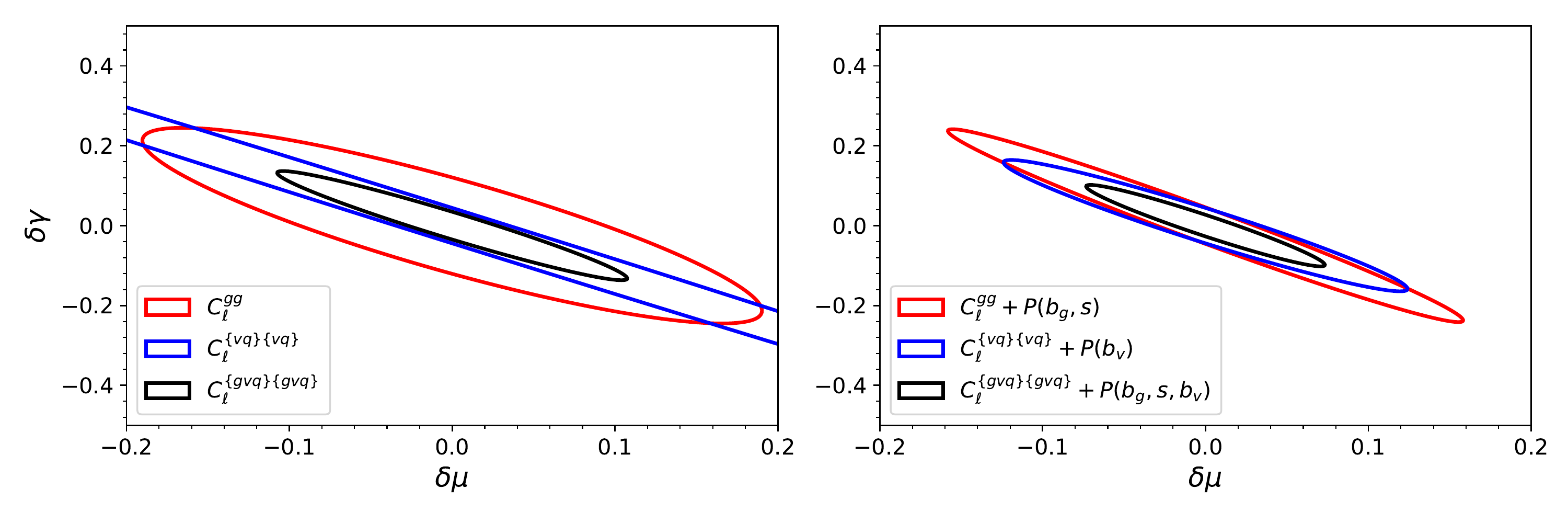}
\caption{\label{fig:ellipse} 1-$\sigma$ contours of forecasted uncertainties of $\delta\mu$ and $\delta\gamma$
using different datasets (LSST and CMB experiment with $\Delta_{\rm T} = 0.1\ \mu$K-arcmin), where in the right panel, we show the forecasts with external prior
information on the galaxy bias $\sigma(b_{\rm g}^i)=0.1 b_{\rm g}^i$ and the optical depth bias $\sigma(b_v^i)=0.1$.}
\end{figure}

In our forecast, we take a conservative cut of $\ell_{\rm max} = 50$ for the angular power spectra of galaxy number counts and moments of the remote dipole and quadrupole fields.
We consider $N_{\rm bins}$ redshift bins with equal width in comoving distance, covering the range $0<z<3$.
In general, the larger $N_{\rm bins}$, the thinner each redshift bin and the larger number of modes available for use.
We use $N_{\rm bins} = 40$ ensuring all redshift bins are wider than the expected redshift error $\sigma_z$ of the LSST gold sample. In addition, all our forecast results
are based on $f_{\rm sky} = 1$, and therefore all the uncertainties obtained should be multiplied by a factor $\sqrt{f_{\rm sky}^{-1}}$ for partial sky coverage.
The results of our forecast are shown in Table \ref{table:forecast} and Fig.~\ref{fig:ellipse}.

\begin{table*}[h]
    \centering
\begin{tabular}{ c| c  c| c c |c c | c  c | c  c}
 \multirow{2}{*}{Dataset} &
 \multicolumn{2}{c|}{$C_\ell^{\rm gg}$} &
 \multicolumn{2}{c|}{$C_\ell^{\rm vv}$} &
 \multicolumn{2}{c|}{$C_\ell^{\rm qq}$} &
 \multicolumn{2}{c|}{$C_\ell^{ \{\rm vq\}\{\rm vq\} }$}   &
 \multicolumn{2}{c}{$C_\ell^{ \{\rm gvq\}\{\rm gvq\} }$}
 \\
  &priors \xmark   &  priors \cmark &  priors \xmark &priors \cmark &  priors \xmark & priors \cmark &  priors \xmark & priors \cmark  &  priors \xmark & priors \cmark \\
    \hline
 $\Delta_{\rm T} = 0.1$  &  \multirow{3}{*}{$(0.19, 0.24)$} & \multirow{3}{*}{$(0.16, 0.24)$}
  & $*$ & $(0.13, 1.50)$ & $(0.97,1.28)$ & $(0.75,1.01)$ & $(0.53, 0.68)$ & $(0.12, 0.16)$ & $(0.11, 0.14)$& $(0.07, 0.10)$\\
 $\Delta_{\rm T} = 1.0$ &  &   & * & $(0.13,1.55)$ &$(5.29,7.03)$ & $(4.76,6.43)$ &$(1.28,1.62)$ & $(0.13,0.23)$ &$(0.12,0.15)$ & $(0.08,0.11)$ \\
 $\Delta_{\rm T} = 5.0$ &  &   & * & $(0.13,1.67)$ & $*$ & $*$ &$(2.80,3.57)$ & $(0.13,0.54)$ &$(0.14,0.18)$ & $(0.08,0.13)$ \\
\end{tabular}
\caption{\label{table:forecast} Forecasted constraints $\sqrt{f_{\rm sky}^{-1}}\sigma(\delta\mu, \delta\gamma)$ from different datasets, where the
LSST gold sample is used for galaxy number counts and the reconstruction noise on the remote dipole/quadrupole fields are based on the LSST gold sample and a CMB experiment with three representative  sensitivities ($\theta_{\rm FWHM}=1.5$ arcmin, $\Delta_{\rm T} = 0.1/ 1.0/ 5.0 \ \mu$K-arcmin). Here we have used notation $*$ for large uncertainties $(>10, >10)$.
For comparison, the constraints from the primary
CMB alone expected from a Planck-like experiment and from an ideal cosmic-variance-limited (CVL) experiment are $\sigma(\delta\mu,\delta\gamma)|_{\rm Planck} = (0.66,1.51)$ and
$\sigma(\delta\mu,\delta\gamma)|_{\rm CVL} = (0.27,0.59)$. }
\end{table*}

As shown in Fig.~\ref{fig:clgg}, the MG parameter $\delta\gamma$ changes
the galaxy angular power spectrum $C_\ell^{\rm gg, i}$ by a nearly
scale-independent factor, which is strongly degenerate with the galaxy bias $b_{\rm g}^i$.
Fortunately, the degeneracy is broken by the power spectra across different redshift bins.
Therefore the constraint of $(\delta\mu, \delta\gamma)$ only marginally improves by adding bias priors $\mathcal P(b_{\rm g}^i) = 0.1 b_{\rm g}^i$ and  $\mathcal P(s^i) = 0.1 s^i$
(see Table \ref{table:forecast} and Fig.~\ref{fig:ellipse}).

As shown in Fig.~\ref{fig:clqv}, the angular power spectrum of the remote dipole field $C_\ell^{\rm vv,i}$
is sensitive to $\delta\mu$, but not $\delta\gamma$. The parameter $\delta\mu$ changes $C_\ell^{\rm vv}$ by a nearly
scale-independent factor, which is strongly degenerate with the optical depth bias $b_v$. Therefore
both $\delta\mu$ and $\delta\gamma$ are unconstrained from kSZ tomography without a prior on the optical depth bias $b_v$.
Imposing optical depth bias priors $\mathcal P(b_v)$ (we take as $\sigma(b_v^i)=0.1$ \cite{Madhavacheril:2019buy}),
we find $\delta\mu$ is constrained by kSZ tomography with uncertainty $\sigma(\delta\mu)\approx 0.13$,
which is almost independent of the CMB experiment sensitivities, i.e,
the constraint on $\delta\mu$ is largely limited by its degeneracy with the optical depth bias. The angular power spectrum of the remote quadrupole field $C_\ell^{\rm qq}$ depends on both parameters $\delta\mu$ and $\delta\gamma$ in a scale-dependent way. The uncertainties of $\delta\mu$ and $\delta\gamma$ constrained from pSZ tomography are largely limited by the reconstruction noise, and imposing optical depth bias $b_v$ priors only slightly reduces the uncertainties. This is evident from the large improvement on the constraints using the remote quadrupole field for CMB experiments with higher sensitivity.

Using both the remote dipole and quadrupole fields $C_\ell^{ \{\rm vq\}\{\rm vq\} }$, the $\delta\mu-b_v$ degeneracy is further broken,
and the uncertainties of $(\delta\mu,\delta\gamma)$ improve by a factor of $\gtrsim 2$ compared with the constraint from the remote quadrupole field only.
Imposing at $10 \%$ prior on the optical depth bias, we again find $\sigma(\delta\mu)\approx 0.13$, independent of the sensitivity of the CMB experiments we considered.

Using the number counts and remote fields $C_\ell^{ \{\rm gvq\}\{\rm gvq\} }$ without any priors, we obtain a better
constraint than that from $C_\ell^{ \{\rm vq\}\{\rm vq\} }+ \mathcal P(b_v)$, especially for the low-sensitivity CMB experiment
we considered. Imposing
both galaxy bias priors $\mathcal P(b_g)$ and optical depth bias priors $\mathcal P(b_v)$ further reduces the uncertainty on
both MG parameters to the $\mathcal{O}(0.1)$ level.

In summary:  although the remote dipole field is sensitive to $\delta\mu$, this parameter is not well
constrained by kSZ tomography alone due to a strong degeneracy between $\delta\mu$ and the optical depth bias $b_v$.
To better constrain $\delta\mu$, we can add prior information  $\mathcal P(b_v)$ from other tracers of the electron distribution
or use larger dataset $C_\ell^{ \{\rm gvq\}\{\rm gvq\} }$ in which the $\delta\mu-b_v$ degeneracy is broken.
In a similar way, $\delta\gamma$ is not well constrained by pSZ tomography $C_\ell^{\rm qq}$ alone
due to a degeneracy with $\delta\mu$, which can be broken using a larger dataset $C_\ell^{ \{\rm vq\}\{\rm vq\} }$
or  $C_\ell^{ \{\rm gvq\}\{\rm gvq\} }$  (see Fig.~\ref{fig:ellipse}).
The constraint from the full dataset $C_\ell^{ \{\rm gvq\}\{\rm gvq\} }$ without any priors is better than
that from either $C_\ell^{\rm gg} + \mathcal P(b_{\rm g}, s)$ or $C_\ell^{ \{\rm vq\}\{\rm vq\} } + \mathcal P(b_v)$.

\section{Conclusions}
\label{sec:summary}
In this paper, we explored the potential contribution of SZ tomography from future cosmological datasets to tests of GR on cosmological scales. The remote
dipole and quadrupole fields reconstructed using SZ tomography are sensitive to modifications of gravity in a number of complimentary ways.
We have chosen as our example of modified gravity a two-parameter $(\delta \mu, \delta \gamma)$ modification of the linearized Einstein equations, where the growth of structure
is impacted in proportion to the relative importance of dark energy in the energy budget. In this parameterization, $\delta \mu$ affects the strength of gravitational clustering while
$\delta \gamma$ encodes the gravitational slip (e.g. the non-equality of the Bardeen potentials). The remote dipole field is sensitive to $\delta \mu$ through the enhancement/weakening of the
peculiar velocity field in deeper/shallower potential wells. Because the ISW contribution to the remote dipole field is far smaller than the Doppler contribution from peculiar velocities,
this observable has limited sensitivity to $\delta \gamma$. The remote quadrupole field is sensitive to both $\delta \mu$ and $\delta \gamma$, primarily due to the significant
contribution of the ISW effect to this observable. Unlike the primary CMB, where the late-time ISW effect makes a significant contribution to a rather limited
number of modes~\cite{Aghanim:2018eyx,Ade:2015rim}, the ISW effect makes an important contribution to the remote quadrupole field everywhere. The power of this
observable is therefore limited by the fidelity of the reconstruction, and not cosmic variance.

We have forecasted the possible constraints on $(\delta \mu, \delta \gamma)$ using a next-generation galaxy survey such as LSST and a high-resolution, low-noise CMB experiment such as CMB-S4.
A major limitation on using kSZ/pSZ tomography for testing gravity is the optical depth degeneracy (our inability to use a tracer of LSS to perfectly infer the distribution of electrons),
which we model as a redshift-dependent multiplicative bias $b_v$ on the remote dipole and quadrupole fields. In the absence of a prior on $b_v$, the remote dipole field cannot be used to constrain
modified gravity due to a large degeneracy between $b_v$ and $\delta \mu$. This degeneracy is not as problematic for the remote quadrupole field. However, due to the large reconstruction noise,
the constraints on modified gravity from the remote quadrupole field are not competitive. Constraints from the galaxy number counts themselves also suffer from a degeneracy
between the galaxy bias and the MG parameters. A major result of this paper is that these degeneracies can be largely mitigated by using correlations between the galaxy number counts and the remote
dipole/quadrupole fields. Comparing with galaxies-only, the uncertainties of $\delta \mu$ and $\delta\gamma$ decrease by
$\sim 40\%$  when including the remote dipole/quadrupole fields, yielding  limits $\sigma(\delta \mu, \delta \gamma)$ of $(0.12,0.15)$ for the median CMB noise considered (assuming data on the full sky). If $10\%$ priors on the galaxy bias and optical depth bias are included, then these constraints can be further improved by $\sim 30 \%$. Further improvement is available as the CMB noise is lowered, which makes more information from the remote dipole/quadrupole fields accessible. This can be compared with the cosmic-variance limited constraint from the primary CMB temperature and polarization of $(0.27,0.59)$.

Although we have made a number of idealistic assumptions, such as data on the full sky and no foregrounds or systematics among others, our result is intended to determine if SZ tomography could in principle be an important tool for testing gravity with cosmology. SZ tomography will be feasible with future cosmological datasets, providing additional information on modifications of gravity for `free'. In this respect, we view our results as encouraging, motivating more detailed analyses with existing and future cosmological datasets.

\section{Acknowledgments}
We would like to thank Niayesh Afshordi, Juan Cayuso, Adrienne Erickcek, James Mertens, and Moritz Munchmeyer for helpful discussions.
This research was supported in part by Perimeter Institute for Theoretical Physics. Research at Perimeter Institute is supported by the Government of Canada through
the Department of Innovation, Science and Economic Development Canada and by the Province of Ontario
through the Ministry of Research, Innovation and Science. MCJ is supported by the National
Science and Engineering Research Council through a Discovery grant.

\appendix
\section{Analytic understanding of perturbations in modified gravity}
\label{sec:analytic}
To obtain some intuition about the effect of MG on various observables, we now analyze the evolution of structure in large scale
small scale limits \cite{Baker:2015bva, Bertschinger2006, Hu2007}.
Without loss of generality, we focus on late-time evolution when radiation energy is far smaller than matter energy and
energy-momentum equation of matter is written as
\begin{subequations}
  \begin{align}
     \Dm' + k_H V & = 3\zeta' \label{eq:continuity} \ \Leftrightarrow Y'+ k_H V = 0, \\
     V' + V & =   k_H\Psi \ , \label{eq:Euler}
  \end{align}
\end{subequations}
where $' = d/d\ln a$, $k = k/aH$, $\zeta = \Phi + V/k_H$ is the gauge-invariant curvature perturbation and $Y = \Dm-3 \zeta$.
For later use, we define $\Hm^2 = 8\pi G\rho_{\rm m}a^3/3$. Assuming the background evolution
in MG is same to that in GR + $\Lambda$CDM, it easy to see
\be\label{eq:LCDM}
2HH' + 3H_{\rm m}^2/a^3 = 0\ .
\ee

Substituting Eq.~(\ref{eq:slip},\ref{eq:continuity}) into Eq.~(\ref{eq:Possion}), we obtain
\be
-\Psi = \frac{3}{2}\frac{\Hm^2}{ak^2}\mu \frac{ Y -3\frac{Y'}{k_H^2} }{1+\frac{9}{2}\frac{\Hm^2}{ak^2}\mu\gamma}\ ,
\ee
Combining the above equation and Eq.~(\ref{eq:Possion}), we find
\be\label{eq:DY}
\Dm = \frac{ Y -3\frac{Y'}{k_H^2} }{1+\frac{9}{2}\frac{\Hm^2}{ak^2}\mu\gamma}\ .
\ee
Differentiating Eq.~(\ref{eq:continuity}) and using Eq.~(\ref{eq:Possion},\ref{eq:Euler}),
we obtain the evolution equation for the overdensity $Y$
\be\label{eq:Yevol}
Y'' + Y'\left(2+\frac{H'}{H}+ \frac{\frac{9}{2}\frac{\Hm^2}{ak^2}\mu}{1+\frac{9}{2}\frac{\Hm^2}{ak^2}\mu\gamma} \right)
-Y\frac{\frac{3}{2}\frac{\Hm^2}{a^3H^2}\mu}{1+\frac{9}{2}\frac{\Hm^2}{ak^2}\mu\gamma} = 0\ .
\ee

In the large scale limit $k\rightarrow0$, Eq.~(\ref{eq:Yevol}) is simplified to
\be
Y'' + Y'\left(2 + \frac{H'}{H}+ \frac{1}{\gamma}\right) -Y\frac{k_H^2}{3}\frac{1}{\gamma} = 0\ ,
\ee
which (along with Eqs.~[\ref{eq:continuity},\ref{eq:DY}]) shows that $Y$,
therefore $V$ and $\Psi+\Phi$ only depends on $\gamma$, while $\Dm$ depends on both $\mu$ and $\gamma$.
In the opposite limit $k\rightarrow\infty$, Eq.~(\ref{eq:Yevol}) is simplified as
\be
Y'' + Y'\left(2+\frac{H'}{H} \right)
-Y \left(\frac{3}{2}\frac{\Hm^2}{a^3H^2}\mu\right)= 0\ ,
\ee
which shows that $Y$, $V$ and $\Dm$ only depend on $\mu$, while $\Psi+\Phi$ depends on both $\mu$ and $\gamma$.

\section{Consistency requirement on super-horizon modes}
\label{sec:consistency}
Assuming that MG is a metric theory in a statistically homogeneous and isotropic cosmology where energy-momentum is
conserved, Bertschinger \cite{Bertschinger2006} showed that the curvature perturbation $\zeta$ remains a constant to leading order,
\be\label{eq:constraint}
\zeta' = \mathcal{O}(k_H^2 \zeta)\ ,
\ee
where
\be\label{eq:zetaprime}
\zeta' = \Phi' + \Psi + \frac{V}{k_H}\frac{H'}{H} \ .
\ee
The consistency between the super-horizon constraint (\ref{eq:constraint})
and the MG parameterization (\ref{eq:Possion},\ref{eq:slip}) along with the energy-momentum
conservation equations (\ref{eq:continuity},\ref{eq:Euler}) has been explicitly examined
in previous works \cite{Zhao:2008bn, Baker:2015bva}. Here we perform an order of magnitude estimate
pointing out a subtle difference in super-horizon evolution in GR versus MG.
Since $\mathcal{O}(V/k_H)=\mathcal{O}(\zeta)$, we can
parameterize Eq.~(\ref{eq:constraint}) as \cite{Hu2007}
\be\label{eq:fzeta}
\lim_{k_H\rightarrow 0} \zeta' = \frac{1}{3}f_\zeta k_H V\,
\ee
where $f_\zeta$ is of $\mathcal{O}(1)$ and needs to be determined by the consistency requirement.
From Eq.~(\ref{eq:Possion}), we obtain
\be
-(\Psi/\mu) - (\Psi/\mu)' = \frac{3}{2}\frac{H_{\rm m}^2}{a} \Dm'\ .
\ee
Combining Eq.~(\ref{eq:continuity}) with Eq.~(\ref{eq:fzeta}),
we have
\be
\lim_{k_H\rightarrow 0} (\Psi/\mu) + (\Psi/\mu)' =\frac{3}{2}\frac{H_{\rm m}^2}{a^3H^2}(1-f_\zeta)\frac{V}{k_H}\ ,
\ee
Combining Eq.~(\ref{eq:constraint},\ref{eq:zetaprime}, \ref{eq:slip}) with the above equation, we obtain
the governing equation of $f_\zeta$
\be
\lim_{k_H\rightarrow 0} \Psi\left(1-\frac{1}{\gamma}-\frac{\gamma'}{\gamma}-\frac{\mu'}{\mu} \right)
-\frac{V}{\gamma k_H}\frac{H'}{H}
= - (1-f_\zeta)\frac{V}{k_H}\frac{H'}{H}\ .
\ee
where we have used Eq.~(\ref{eq:LCDM}). It is easy to see $f_\zeta|_{\rm GR}= 0$, while $f_\zeta|_{\rm MG}\neq 0$, i.e. ,
\be
\lim_{k_H\rightarrow 0}\zeta'|_{\rm GR} = \mathcal{O}(k_H^3 \zeta)\ , \quad  \lim_{k_H\rightarrow 0}\zeta'|_{\rm MG} = \mathcal{O}(k_H^2 \zeta)\ .
\ee
This explains why it is safe to set $\zeta' =0$ in Eq.~(\ref{eq:continuity}) in analyzing super-horizon evolution in GR, while the same approximation
leads to a wrong solution in MG.

\bibliography{references}

\end{document}